# Line geometry and electromagnetism II: wave motion

D. H. Delphenich
Kettering, OH 45440

**Abstract.** The fundamental role of line geometry in the study of wave motion is first introduced in the general context by way of the tangent planes to the instantaneous wave surfaces, in which it is first observed that the possible frequency-wave number 1-forms are typically constrained by a dispersion law that is derived from a constitutive law by way of the field equations. After a general review of the basic concepts that relate to quadratic line complexes, these geometric notions are applied to the study of electromagnetic waves, in particular.

**Table of contents**

**1. Introduction.** Although one generally thinks of Maxwell's formulation of the laws of electromagnetism as the point in history at which electromagnetic waves first became possible in theory, nonetheless, the concept of light, in particular, as a wave-like motion dates back to Christiaan Huygens in 1690 [**1**]. Sadly, his theory − which was based upon a mechanical analogy with sound waves, but still predicted the speed of propagation to within about 15% accuracy − was immediately overshadowed by Newton's corpuscular theory, which appeared about ten years later. As a result, the further research into the wave theory of light did not begin in earnest for roughly another century with Thomas Young (1803) and Augustin Fresnel (1819).

    In particular, the theory of Fresnel (see, e.g., [**2, 3**]) was still based upon a mechanical conception of the "ether," although he did at least restrict the vibrations that contributed to the wave motion to be transverse ones. As a result of his analysis, Fresnel defined two surfaces in projective space and its dual space that are typically called the *ray surface* and the *wave surface* (or *normal surface*), respectively. They represent dispersion laws for (linear, dispersionless) optical media, which basically constrain the possible frequency-wave number 1-forms $k$ that can be associated with electromagnetic waves. Because they

are quartic in the relevant variables, they can lead to such fascinating optical phenomena as birefringence – or double refraction – and conical refraction, which can be either internal or external. Interestingly, although the mechanical basis for the analysis is no longer recognized, if one simply replaces the mechanical stress tensor with the dielectric tensor, the resulting quartic surfaces are still just as useful to this day.

The surfaces that Fresnel defined spawned a considerable amount of purely geometric investigations, and in the Nineteenth Century the reigning school of geometry was projective geometry. It was in the second half of that century when Julius Plücker first defined the foundations of line geometry, which used the concept of a line in $\mathbb{R}P^3$ as its fundamental object. The better part of his seminal work on the subject [**4a**] was devoted to the study of line complexes ([1]), which amount to algebraic sets in the space of all lines in $\mathbb{R}P^3$ that are defined by sets of functions on that space.

In particular, most of his analysis was focused upon quadratic line complexes, which are defined by the zero locus of a single quadratic function on the space of lines in $\mathbb{R}P^3$. Since a line can be represented by the join of two distinct points (Plücker called these lines "rays") or the meet of two distinct planes (he then called these lines "axes"), there are also two way of representing a line complex. The degree of the homogeneous function on rays that defines the complex is its *order* and the degree of the corresponding function on axes is called its *class*.

He defined the notion of the "singularity surface" of a quadratic line complex, which consists of all points in $\mathbb{R}P^3$ at which the "complex cone" of that complex degenerates into a pair of intersecting planes; that is the quadratic equation that defines the cone factors into a pair of linear equations. Dually, the lines of the complex that lie in a fixed plane will envelop of curve, and in the degenerate case that curve will become a pair of points.

One of the most definitive studies of linear and quadratic complexes was the doctoral dissertation [**5a**] of Ernst Kummer (under Heinrich Scherck) ([2]). In it, he examined the geometry of a particular class of singularity surfaces that are associated with line complexes or order and class two, namely, the ones that have 16 singular points and 16 singular tangent planes. In the former case, the word "singular" now refers to the points at which the tangent plane degenerates into a tangent cone, while in the latter case, it refers to tangent planes that contact the surface at more than one point, and thus envelop a singular cone. Furthermore, the 16 points exhibit a high degree of symmetry as the orbit of the action of a finite group with sixteen elements.

Such singularity surfaces came to be called "Kummer surfaces" (see [**6**] for an exhaustive discussion from the classical era), and went through various subsequent generalizations, namely, to K3 surfaces (named for Kummer, Kähler, and Kodaira, as well as the mountain peak K2) and Calabi-Yau manifolds. The Fresnel wave surface becomes a degenerate case of a Kummer surface in which the 16 points group themselves on the faces of a tetrahedron with four on each face.

---

([1]) For Plücker's comments on the role of line complexes in geometrical optics, see [**4b**].
([2]) See Kummer's comments on geometrical optics, as well [**5b, c**].

What was missing from most of the definitive classical references on line complexes was the Plücker-Klein embedding of the lines in $\mathbb{R}P^3$ into the vector spaces $\Lambda_2$ and $\Lambda^2$ of bivectors and 2-forms over $\mathbb{R}^4$ as lines through the origin that contain only decomposable bivectors and 2-forms. Since this topic was discussed in Part I of this series of articles, we shall assume that it is known to anyone who is reading this installment. The reason that the Plücker-Klein embedding is missing from line geometry (although it is contained in it implicitly) is that the leading figures of line geometry in its earliest manifestations – in particular, Klein himself – were apparently unimpressed by Grassmann's theory of exterior algebra (*Ausdehnungslehre* = theory of extensions), and it was not until Burali-Forti [7] began to apply Grassmann's exterior algebra to projective and differential geometry that this bridge became very well established.

Since the modern formulation of the theory of electromagnetism, whether in its traditional Lorentz-invariant metric form or its pre-metric form [8, 9] – makes fundamental use of Grassmann's exterior algebra, when one asks the question "What is the 'pre-metric' geometry of spacetime?," one finds that a reasonable response might be "projective geometry, and especially line geometry." Thus, this series of articles has been charged with the responsibility of exploring some of the applications of line geometry in the theory of electromagnetism by exploiting the bridge between them that the Plücker-Klein embedding defines.

In Part I of the series [10], the emphasis was on establishing the elementary concepts of line geometry and their application to the fields that enter into the theory of electromagnetism. In this part, we shall focus more specifically on the application of the theory of quadratic line complexes to the study of certain special classes of electromagnetic constitutive laws, namely, linear, dispersionless ones. In particular, we shall show how the derivation of the dispersion law for plane waves in such media runs parallel to the derivation of the singularity surface that is associated with the quadratic complex that the constitutive law defines.

Before we commence with the applications to electromagnetism, we first point out how wave motion in three-dimensional space typically defines a fundamental decomposable 2-form $dt \wedge d\phi$ that annihilates the tangent spaces to the instantaneous wave surfaces. Since these surfaces are ultimately the intersections of isophase hypersurfaces, which are level surfaces of the phase function $\phi$, and level surfaces of the time coordinate $t$, the tangent planes in question will be intersections of the tangent hyperplanes to those hypersurfaces. When one goes to the projectivized tangent bundle, the hyperplanes become (projective) planes that intersect in a (projective) line that is associated with $dt \wedge d\phi$ by way of the dual Plücker-Klein embedding. The introductory section then concludes by discussing the role of dispersion laws for wave motion, in general.

In the following sections, we specialize the general discussion of wave motion by introducing electromagnetic constitutive laws, and then showing how some of them can relate to quadratic line complexes, and showing how the process of deriving a singularity surface or dispersion law works. Particular attention is given to the most elementary – if not purely geometric – constitutive law that is associated with the Hodge * isomorphism when one has a Lorentzian metric defined to begin with. One sees that, in fact, if one starts with such an isomorphism abstractly then the fact that it defines an "almost-complex" structure on the bundle of 2-forms allows one to define a conformal class of Lorentzian metrics – i.e., the dispersion law for light cones – in a more direct manner than the usual construction of the singularity surface.

**2. The representation of wave motion in space.** One finds that many of the constructions that are so intrinsic to the role of line geometry of electromagnetism are closely-related to the modeling of wave motion in space ([1]). That is based upon simply the fact that waves in space involve the motion of surfaces, and the tangent planes to surfaces can be represented by bivector fields or 2-forms, when one adds the time dimension to the three dimensions of space. Of course, a plane through the origin in $\mathbb{R}^4$ will become a line in $\mathbb{R}P^3$.

*a. Phase structures on space.* We start by defining a *phase structure* on space (which will be simply $\mathbb{R}^3$, for now) to be simply a codimension-one foliation of space by the *isophase* surfaces of some differentiable function $\phi : \mathbb{R}^3 \to \mathbb{R}$; i.e., the level surfaces of the function. Its differential then defines the *wave-number* 1-form:

$$k_s = d\phi = k_i \, dx^i, \qquad k_i = \frac{\partial \phi}{\partial x^i}, \qquad (2.1)$$

although this definition can be generalized according to the degree of integrability of $k$. The tangent planes to the isophase surfaces are then the annihilating planes of the wave-number 1-form $k$. Note that, in principle, the same construction will apply to an electric potential function $f$ with $E = -d\phi$. The annihilating planes of $d\phi$ will then be the equipotential surfaces.

This way of defining the isophase surfaces amounts to the *envelope* construction; that is, one could say they are the envelopes of their tangent planes. One can also define them as *loci* by way of embeddings $\sigma : \Sigma \to \mathbb{R}^3$, $(u, v) \mapsto \sigma^i(u, v)$, where $\Sigma$ is some subset of $\mathbb{R}^2$ (plane, cylinder, sphere, ellipsoid, etc.), and $\sigma$ is compatible with $\phi$ in the sense that $\sigma(\Sigma)$ is a level set of $\phi$:

$$\phi \cdot \sigma = \text{const.} \qquad (2.2)$$

Since $\sigma$ is assumed to be an embedding, the differential $d\sigma$ is assumed to have rank two, and therefore the image of $T_{(u, v)}\Sigma$ under $d\sigma$ will be a plane in $T_{\sigma(u, v)}\mathbb{R}^3$, which is, moreover, the annihilating plane of $k$ at that point, since differentiating (2.2) will give:

$$d\phi \cdot d\sigma = 0. \qquad (2.3)$$

---

([1]) One might also confer the author's study of the pre-metric foundations of wave mechanics [**11**].

Furthermore, since $d\sigma$ has rank two the image $\{\mathbf{e}_u = \sigma_*\partial_u, \mathbf{e}_v = \sigma_*\partial_v\}$ of the canonical frame $\{\partial_u, \partial_v\}$ on $\mathbb{R}^2$ will be a 2-frame in $T_{\sigma(u, v)}\mathbb{R}^3$ at each point $\sigma(u, v)$. Thus, one can define a bivector field $\mathbf{e}_u \wedge \mathbf{e}_v$ on the image of $\sigma$.

It should be pointed out that typically the embedding $\sigma$ takes the form of an integral submanifold for a differential system that amounts to a choice of plane in each tangent space. Hence, to that extent that integration usually follows the formulation of differential systems in terms of first principles, one should probably regard the envelope construction as more fundamental than the locus construction, *a priori*.

So far, this picture is only useful for time-invariant geometrical optics, which is the optical analogue of steady fluid flow or static electric fields in space. We can now introduce a time dimension into the picture in various ways, and the one that is most convenient to the current discussion is that of extending $\mathbb{R}^3$ to $\mathbb{R}^4$ by way of a time coordinate $t = x^0$, which also extends the phase function $\phi$ to the pair of functions $(t, \phi)$: $\mathbb{R}^4 \to \mathbb{R}^2$, $x \mapsto (t(x), \phi(x))$. One then defines the *frequency-wave number 1-form:*

$$k = d\phi = k_\mu\, dx^\mu, \qquad \mu = 0, \ldots, 3, \qquad k_\mu = \frac{\partial \phi}{\partial x^\mu}, \qquad (2.4)$$

in which we have defined:

$$k_0 = \omega. \qquad (2.5)$$

Although it is commonplace to include a factor of $1/c$ and $c$ in these expressions, in order to make the coordinates have the same units, when one is dealing with the propagation of general waves, it is not always meaningful to refer to the speed of propagation as a single constant. In general, it does not have to be constant or even single, as one would find in inhomogeneous, anisotropic media.

The envelope construction then looks at the intersection of the two hypersurfaces that are defined by the two functions $t$ and $\phi$ – i.e., the intersection of the isophase hypersurfaces with the *isochrones*, which are the level hypersurfaces of $t$. We shall call these surfaces the *instantaneous wave surfaces*, since they represent the isophase surfaces for each point in time.

Furthermore, in order to make the instantaneous wave surfaces be actual surfaces in every case, we assume that the intersection of the level hypersurfaces is transversal, so basically $dt$ will not be collinear with $d\phi$ at any point of $\mathbb{R}^4$. Hence, one can define a non-zero 2-form by way of $dt \wedge d\phi$. The annihilating planes of this 2-form in the tangent spaces to $\mathbb{R}^4$ will be the tangent planes to the instantaneous wave surfaces; i.e., they will define the differential system whose integral submanifolds are the instantaneous wave surfaces.

We can now extend the embedding $\sigma$ to an embedding $\sigma: \Sigma \to \mathbb{R}^4$, which will have the novel aspect that the time coordinate $t$ will become a function of $(u, v)$. However, it does have useful property that since $\sigma$ is still assumed to be an embedding, $d\sigma$ will still be of rank two at every point of $\Sigma$, and we can still speak of the bivector field $\mathbf{e}_u \wedge \mathbf{e}_v$ on the image of $\sigma$. Furthermore, the tangent subspace that is spanned by $\mathbf{e}_u$, $\mathbf{e}_v$ will be the annihilating plane of $dt \wedge d\phi$. Hence:

$$(dt \wedge d\phi)(\mathbf{e}_u \wedge \mathbf{e}_v) = 0, \tag{2.6}$$

and the bivector field $\mathbf{e}_u \wedge \mathbf{e}_v$ is dual to the 2-form $dt \wedge d\phi$ under the bilinear pairing of 2-forms and bivector fields.

One can then think of the bivector $\mathbf{e}_u \wedge \mathbf{e}_v$ as something that comes from the integration of the exterior differential system:

$$dt \wedge d\phi = 0. \tag{2.7}$$

One might generalize this picture somewhat by starting with more general 1-forms $\tau$ and $k$, in place of $dt$ and $d\phi$, respectively, and saying that the exterior differential system in question is defined by the intersection of the annihilating 3-planes of the 1-forms $k$ and $\tau$.

One must note that if the tangent planes to the instantaneous wave surfaces are the fundamental objects then there will be nothing unique about either $dt$ or $d\phi$. In particular, since the time coordinate is not generally defined universally in the context of relativistic physics, one could just as well use any other 1-form $a = \alpha\, dt + \beta\, d\phi$ in the plane spanned by $dt$ and $d\phi$, except, of course, $d\phi$ itself. Such an $a$ will differ from $dt$ by a rotation in that plane and a rescaling, so when one goes to the point $[a]$ in $\mathbb{R}P^{3*}$, the only difference will be the rotation. Hence, the choice of $a$ becomes essentially a choice of "gauge."

The way that this picture relates to line geometry in $\mathbb{R}P^3$ comes from the Plücker-Klein embedding (cf., Part I), which associates each line $[a, b]$ in $\mathbb{R}P^3$ with an equivalence class $[\mathbf{a} \wedge \mathbf{b}] \in P\Lambda_2$ ($\equiv P\Lambda_2 \mathbb{R}^4$) of scalar multiples of a decomposable bivector $\mathbf{a} \wedge \mathbf{b} \in \Lambda_2$ ($\equiv \Lambda_2 \mathbb{R}^4$) where the vectors $\mathbf{a}, \mathbf{b} \in \mathbb{R}^4$ cover the points $a$, $b$, resp., under the projection $\mathbb{R}^4 - 0 \to \mathbb{R}P^3$; i.e., $a = [\mathbf{a}]$, $b = [\mathbf{b}]$. This is what Plücker refers to as the representation of a line in $\mathbb{R}P^3$ as a *ray*; i.e., the join of two distinct points. Dually, since the plane of $[\mathbf{a} \wedge \mathbf{b}]$ can also be annihilated by a unique equivalence class $[\Omega]$ of scalar multiples of some decomposable 2-form $\Omega = \alpha \wedge \beta$ (as long as space is four-dimensional), one can associate the line $[a, b]$ with $[\Omega]$. This is what Plücker refers to as the representation of a line in $\mathbb{R}P^3$ as an *axis*; i.e., the meet of two distinct planes, namely, the annihilating planes of $[\alpha]$ and $[\beta]$.

Hence, one can think of the fundamental object for wave motion in space as representing either a plane through the origin in $\mathbb{R}^4$, a point on the Klein quadric, or a line in $\mathbb{RP}^3$ for each point of the spacetime manifold. (If one prefers to start with a general differentiable manifold for spacetime then $\mathbb{R}^4$ and $\mathbb{RP}^3$ will pertain to the tangent spaces and their projectivizations; viz., the sets of all lines through their origins.)

Since we are starting with a decomposable 2-form – viz., $dt \wedge d\phi$ – as our fundamental object, it is, of course, more natural to think of the tangent planes to the instantaneous isophase surfaces as also being represented in $\mathbb{RP}^3$ by the line of intersection of the planes that are defined by $[dt]$ and $[d\phi]$. Hence, in the language of Plücker we are representing that line as an axis; by contrast, the bivector $\mathbf{e}_u \wedge \mathbf{e}_v$ represents the same line as a ray.

One can treat $k$ as if its components were the homogeneous coordinate for a coordinate system on $\mathbb{RP}^{3*}$. The corresponding inhomogeneous coordinates:

$$K_i = \frac{k_i}{\omega} \tag{2.8}$$

are basically the inverses of phase velocities, which can also be regarded as indices of refraction in each spatial direction. This tends to suggest that the usual definition of phase velocity is not as geometrically natural as that of index of refraction. The coordinates $K_i$ also basically represent the direction to a normal to the wave front that is described by the isophase surfaces, since they are clearly proportional to the components of the covector $k_i$.

*b. Dispersion laws for waves.* So far, the geometric picture of wave motion that we have defined is lacking one fundamental piece in the form of a *dispersion law*. Such a law typically takes the form of a constraint that the 1-form $k = d\phi$ must lie on some specified hypersurface in each cotangent space:

$$P[k] = \begin{cases} 0, \\ k_0^d, \end{cases} \tag{2.9}$$

in which $P$ is a homogeneous function of degree $d$.

The case in which the right-hand side is zero is most commonly associated with massless waves, while the non-zero case can suggest massive waves, such as the ones that are described by the Klein-Gordon equation, or ones that propagate in certain types of plasmas.

Since $P[k]$ is homogeneous, one can also represent the dispersion law as a surface in $\mathbb{RP}^3$ that is typically associated with an inhomogeneous algebraic equation of the form:

$$P[K] = K_0^d. \tag{2.10}$$

Typically, the way that one gets to a dispersion law in theory is by starting with the equations for the field in question – whether that be a field of mechanical strains or displacements or a field of electromagnetic displacements, for example – and assuming that a "wave-like solution" takes a certain form that involves a frequency-wave-number 1-form *k*, such as a plane wave or the generalizations of geometrical optics, and then seeing what sort of algebraic relationship results for *k* upon substituting the general solution into the field equations. Of course, this process is usually more useful in the context of linear wave motion, since in the nonlinear case different forms of wave-like solutions can produce distinct dispersion laws; i.e., a dispersion law is generally peculiar to a specific form for a wave-like solution.

One can also treat a dispersion law as a purely empirical relationship that one derives by experimentally passing waves of some type through a medium that supports the propagation of waves and directly measuring how the frequency of the wave relates to the wave number, if not other parameters, such as the polarization state of the wave or its amplitude.

In either event, one will find that the dispersion law for the propagation of waves in a medium is intimately linked to the constitutive laws for the medium, whether mechanical or electromagnetic. We shall see that for linear, dispersionless ([1]), electromagnetic media the path from the constitutive law to the dispersion law is established by purely line-geometric constructions, namely, the construction of the *singularity surface* (or *focal surface*) of the quadratic line complex that the constitutive law defines. Typically, that surface will be defined by a homogeneous quartic function of the frequency-wave number 1-form, such as the Fresnel surface that one encounters in the optics of biaxial, anisotropic, electromagnetic media.

Since the geometry of wave motion in space seems to start off as a fundamental 2-form $\Omega$ – or really, the class $[\Omega]$ of all non-zero scalar multiples – which only includes the 1-form *k* as one of its exterior factors, one might wonder if the dispersion law itself is related to some corresponding condition on the fundamental 2-form, if not derived from it. Indeed, this is typically the case, but one shows that by a rather circuitous process that starts off with the constitutive law for the medium, whether mechanical or electromagnetic. In a later section, we shall treat the most common way that a condition on the fundamental 2-form can imply a dispersion law, which is the one that relates to the conformal Lorentzian dispersion law.

**3. Quadratic line complexes.** If $\mathcal{L}(3)$ represents the set of all lines in $\mathbb{R}P^3$ (see Part I [**10**]) then a *line complex* in $\mathbb{R}P^3$ will be generally defined to be a set of lines $[x_0, x]$ that is the level set of some real function $\mathcal{C}: \mathcal{L}(3) \to \mathbb{R}$, $[x_0, x] \mapsto \mathcal{C}[x_0, x]$. When the line $[x_0, x]$ is represented as a point $[\mathbf{x}_0 \wedge \mathbf{x}]$ on the Klein quadric $\mathcal{K}$, one typically defines $\mathcal{C}$ by way of a homogeneous function $C : \Lambda_2 \to \mathbb{R}$, $\mathbf{a} \wedge \mathbf{b} \mapsto C(\mathbf{a} \wedge \mathbf{b})$. The *degree* of the complex $\mathcal{C}$ is then the degree of the function $C$. When one passes from $\Lambda_2$ to $P\Lambda_2$, since $C$ is homogeneous, it will define a corresponding function $[C] : P\Lambda_2 \to \mathbb{R}$, $[\mathbf{a} \wedge \mathbf{b}] \mapsto [C][\mathbf{a} \wedge \mathbf{b}]$, and if the 4-vectors $\mathbf{a}$ and $\mathbf{b}$ represent the points $a = [\mathbf{a}]$ and $b = [\mathbf{b}]$ in $\mathbb{R}P^3$, resp., then one will have:

$$\mathcal{C}[a, b] = [C][\mathbf{a} \wedge \mathbf{b}] . \tag{3.1}$$

---

[1] This usage of the word "dispersion" refers to the character of the electromagnetic constitutive law itself – viz., the fact that it does not depend upon frequency or wave number – and is distinct form the one that relates to the propagation of waves directly, which specifies how frequency and wave number get coupled to each other.

Although we do not have that $[C][\mathbf{a} \wedge \mathbf{b}] = C(\mathbf{a} \wedge \mathbf{b})$, we do have that the one expression will vanish iff the other one does.

Since the space $\mathcal{L}(3)$ is four-dimensional, as long as a line complex is defined by a single function, the level hypersurfaces will generally be three-dimensional.

*a. Generalities.* A *quadratic line complex* on $\mathbb{R}P^3$ [4, 12] is then defined when the function $C$ is homogeneous quadratic. If $\mathbf{A} = A^I \mathbf{E}_I \in \Lambda_2$ then we can represent the quadratic form of $C$ by:

$$C[\mathbf{A}] = C_{IJ} A^I A^J, \qquad C_{IJ} = C_{JI} . \tag{3.2}$$

In particular, if $x_0, x \in \mathbb{R}P^3$ are two (finite) points whose inhomogeneous coordinates are $X_0^i$, $X^i$ then if we associate them with the homogeneous coordinates $(1, X_0^i)$, $(1, X^i)$ then the line $[x_0, x]$ will be represented by the bivector:

$$\mathbf{x}_0 \wedge \mathbf{x} = \mathbf{e}_0 \wedge (X^i - X_0^i) + \tfrac{1}{2}(X^i X_0^j - X^j X_0^i)\mathbf{e}_i \wedge \mathbf{e}_j , \tag{3.3}$$

and the components $A^I$ of a bivector that represents the line $[x_0, x]$ will be:

$$A^i = X^i - X_0^i, \qquad A^{i+3} = \tfrac{1}{2}\varepsilon^{ijk}(X^j X_0^k - X^k X_0^j) . \tag{3.4}$$

It will be more convenient in what follows to treat $\mathbf{x}$ as $\mathbf{x}_0 + \Delta\mathbf{x}$, with the obvious definition for $\Delta\mathbf{x}$. One will then have:

$$\mathbf{x}_0 \wedge \mathbf{x} = \mathbf{x}_0 \wedge \Delta\mathbf{x} = \mathbf{e}_0 \wedge \Delta X^i \mathbf{e}_i + \tfrac{1}{2}(\Delta X^i X_0^j - \Delta X^j X_0^i)\mathbf{e}_i \wedge \mathbf{e}_j , \tag{3.5}$$

and:

$$A^i = \Delta X^i, \qquad A^{i+3} = \tfrac{1}{2}\varepsilon^{ijk}(\Delta X^j X_0^k - \Delta X^k X_0^j) . \tag{3.6}$$

One can then represent $C$ as a homogeneous quadratic function $C(X_0^i, \Delta X^i)$. If one fixes $x_0$ then one can also define a homogeneous quadratic function of $x$:

$$\mathcal{C}(x_0)(x) = \mathcal{C}[x_0, x], \tag{3.7}$$

which is associated with:

$$C(X_0^j)(\Delta X^i) = C(X_0^j)_{ij} \Delta X^i \Delta X^j, \qquad C(X_0^j)_{ij} = C(X_0^j)_{ji}. \tag{3.8}$$

The coefficients $C(X_0^j)_{ij}$ of this quadratic form are then homogeneous quadratic functions of $X_0^j$.

If $x_0$ is a point in $\mathbb{R}P^3$ then the set $\mathcal{C}(x_0)$ of all $x$ that make $\mathcal{C}(x_0)(x)$ vanish will define a quadratic cone in $\mathbb{R}P^3$ with $x_0$ for its vertex. One refers to such a cone $\mathcal{C}(x_0)$ as a *complex cone* for the complex $\mathcal{C}$. Since a line of the complex goes through every point $x_0$ of $\mathbb{R}P^3$, every point of $\mathbb{R}P^3$ will associated with a complex cone. The lines of the cones will then satisfy the equation:

$$C(X_0^j)_{ij} \Delta X^i \Delta X^j = 0. \tag{3.9}$$

For some points $x_0$, the complex cone $\mathcal{C}(x_0)$ can degenerate to a pair of intersecting planes $[\alpha]$, $[\beta]$. Such a point is then called a *singular* point of the complex. Thus, if $\alpha$ and $\beta$ are 1-forms that represent these planes, one will be able to represent the quadratic form $\mathcal{C}(x_0)$ as:

$$C(X_0)(X) = \alpha(X_0)(X)\, \beta(X_0)(X), \tag{3.10}$$

which will correspond to:

$$C(X_0^j)_{ij} = \tfrac{1}{2}[\alpha_i(X_0)\beta_j(X_0) + \alpha_j(X_0)\beta_i(X_0)]. \tag{3.11}$$

We can define a linear map $c : \mathbb{R}^4 \to \mathbb{R}^{4*}$, $\mathbf{v} \mapsto \tfrac{1}{2}[\alpha(\mathbf{v})\beta + \beta(\mathbf{v})\alpha]$, which will vanish for all $\mathbf{v}$ such that $\alpha(\mathbf{v}) = \beta(\mathbf{v}) = 0$, which is then the line of intersection of the planes defined by $\alpha$ and $\beta$. Hence, the dimension of the kernel of $c$ is greater than 0, so $c$ cannot be invertible ([1]), and its determinant must vanish for any bases chosen on $\mathbb{R}^4$ and $\mathbb{R}^{4*}$.

Therefore, a necessary condition for the factorizability of the quadric that is defined by $C(X_0)$ is the vanishing of the determinant of $C(X_0^j)_{ij}$. That is, the set of all points $x_0$ for which that condition is true is then a hypersurface in $\mathbb{R}P^3$ that one calls the *singularity surface* – or *focal surface* (German: *Brennfläche*) – of the complex $\mathcal{C}$, and which has the equation:

$$|C(X_0^j)_{ij}| = 0. \tag{3.12}$$

---

([1]) One also sees that $c$ is not surjective, since its image is spanned by $\alpha$ and $\beta$.

Although the fact that this matrix is 3×3 and its elements are quadratic functions suggests that one should be dealing with a sextic polynomial, in fact, it will reduce to a quartic upon performing the calculations.

Dually, since a line in $\mathbb{R}P^3$ can also be expressed as the intersection of two distinct planes, which will define a point on the dual Klein quadric $\mathcal{K}^*$ in $P\Lambda^2$, one can define a quadratic line complex by means of a homogeneous quadratic function $C^*$ on $\Lambda^2$. One then says that such a complex has *class two*. If one fixes one of the planes then the complex will define a curve in that plane that is enveloped by all of the lines of the complex that lie in that fixed plane. This curve in a plane is then the dual construction to the complex cone through a point. In the degenerate case, the curve becomes a pair of two points in that plane, which is then a *singular plane*, and the corresponding algebraic condition for that is the vanishing of another homogeneous quartic function on the 1-forms. Although it is not clear at this point, the quartic hypersurface that is defined in this dual construction is actually the same as the one that was defined above. (This fact is asserted by Plücker [**4a**], no. **320** and elaborated upon by Pasch [**13**].)

Further degeneracies can arise, such as when the two planes at a singular point coincide or the two points in a singular plane can coincide. One then refers to such planes and points as *double planes* and *double points*, respectively.

*b. Kummer surfaces.* The points of a singularity surface can be singular, in a different sense, namely, in the sense that the tangent lines to the surface through that point do not define a plane pencil of lines, but, in fact, a quadric cone of lines.

At a non-singular point $x \in \mathbb{R}P^3$, the tangent plane to a surface in $\mathbb{R}P^3$ that is defined by a level point of a differentiable function $F : \mathbb{R}P^3 \to \mathbb{R}$ consists of all tangent vectors **v** $\in T_x\mathbb{R}P^3$ such that:

$$0 = dF|_x(\mathbf{v}) = \frac{\partial F}{\partial X^i}(x)v^i, \tag{3.13}$$

in which the local expression is given with respect to an inhomogeneous coordinate chart about *x* and its natural frame field.

At a singular point, one then has $dF|_x = 0$, or locally:

$$\frac{\partial F}{\partial X^i}(x) = 0, \quad i = 1, 2, 3, \tag{3.14}$$

so equation (3.13) becomes ill-defined, since it will admit all tangent vectors **v** as solutions.

However, not all **v** in $T_x\mathbb{RP}^3$ are, in fact, tangent to the level surface of $F$ through $x$, and to find the ones that are, one must go to the next derivative of $F$ and pull $F$ back to a homogeneous function $\bar{F}: \mathbb{R}^4 - 0 \to \mathbb{R}$, along the projection $\mathbb{R}^4 - 0 \to \mathbb{RP}^3$, namely:

$$\bar{F}(\mathbf{x}) = F[\mathbf{x}]. \tag{3.15}$$

By iterating Euler's theorem for a function that is homogeneous of degree $r$, namely:

$$\frac{\partial \bar{F}}{\partial v^\mu}(x) v^\mu = r\bar{F}(x), \tag{3.16}$$

taking account of the fact that the derivatives of $\bar{F}$ will be homogeneous of degree $r - 1$, one gets:

$$\frac{\partial^2 \bar{F}}{\partial v^\mu \partial v^\nu}(x) v^\mu v^\nu = (r-1)\frac{\partial \bar{F}}{\partial v^\mu}(x) v^\mu. \tag{3.17}$$

Hence, we can replace the condition (3.14) for tangent vectors to level surfaces with:

$$\frac{\partial^2 F}{\partial X^i \partial X^j}(x) v^i v^j = 0, \tag{3.18}$$

after converting back to inhomogeneous coordinates.

The fact that this defines a quadric cone at $x$ is clear since the condition is homogeneous and quadratic in the $v^i$.

The dual condition to saying that $F$ has a singular point is to say that $F^*$ should have a *singular plane* $\pi$, which then makes $dF^*|_\pi = 0$. In fact, a singular point will be incident on a singular plane and the tangent cone to a singular point will be enveloped by an enveloping cone of planes, which will, however, be sextic, not quadric.

A quartic function on $\mathbb{RP}^3$ will have at most 16 singular points and 16 singular planes, and when one attains this maximum number, one will have defined a *Kummer surface* ([1]), at least in the event that that the singular points are all isolated. One further requires that the singular points and singular planes be arranged in what one calls a "$16_6$ configuration," namely, six singular points are incident on each singular plane and six singular planes pass through each singular point.

Due to their $16_6$ configuration, the singular points of a Kummer surface exhibit a high degree of symmetry. Namely, they are an orbit of the action of a group of 16 elements that acts effectively on $\mathbb{RP}^3$ by collineations and consists of various permutations and sign changes of the homogeneous coordinates.

Note that not every Kummer surface is associated with a unique quadratic complex, but generally with a singly-infinite family of them.

---

([1]) Kummer first defined them in his doctoral dissertation [**5a**] (see also his comments in [**5b**]), and they attracted the attention of numerous other geometers of the era (e.g., [**13, 14**]).

**4. Electromagnetic waves.** Perhaps the most profound consequence of Maxwell's equations was the fact that they admitted "wave-like" solutions. Of course, the very definition of the word "wave-like" is open to debate, but in the case of linear field equations, it is sufficient to consider plane electromagnetic waves. One way of testing the field equations to see if they admit wave-like solutions is to assume such a solution that includes a phase function $\phi$, apply the field equations to it, and look at the resulting algebraic equations that relate to the frequency-wave number 1-form $k$ – i.e., the dispersion law for that type of wave. If the equations admit a non-trivial solution then the field equations will admit non-trivial waves.

As a result, one defines an important association of the constitutive law that defines the field equations and the dispersion law that describes the way that electromagnetic waves can propagate in the media. Indeed, this association was being explored in the name of projective geometry even while Maxwell was developing his own theory of electromagnetism, although the geometers were basing their analysis upon the work of Fresnel on optics much earlier in the Nineteenth Century.

Hence, in this section, we shall explore both the physical and the geometrical aspects of how one gets from an electromagnetic constitutive law to a dispersion law for electromagnetic waves. It is in that process that one also finds how the Lorentzian metric on spacetime that accounts for the presence of gravitation can "emerge" from the electromagnetic structure as a possible degenerate case of a dispersion law.

*a. Electromagnetic constitutive laws.* In the metric formulation of Maxwell's equations, the only place in which the Lorentzian metric on the spacetime manifold plays a role is in the definition of the Hodge * isomorphism, which takes 2-forms to bivector fields. However, in the pre-metric formulation of electromagnetism [**8, 9**], the role that is otherwise played by * is played by the composition $\kappa = \# \cdot C$ of two isomorphisms $C : \Lambda^2(M) \to \Lambda_2(M)$ and $\# : \Lambda_2(M) \to \Lambda^2(M)$.

The second of the two isomorphisms is simply the Poincaré duality isomorphism that comes from assuming that the tangent bundle $T(M)$ to the spacetime manifold $M$ is orientable and that a choice of unit-volume element $V \in \Lambda^4(M)$ has been made. The isomorphism is then the one that takes a bivector **A** in any fiber of $\Lambda_2(M)$ to the 2-form $\#\mathbf{A} = i_\mathbf{A} V$.

The first of these maps is the electromagnetic constitutive law, which takes the electromagnetic field strength 2-form $F$ to the electromagnetic excitation ([1]) bivector field $\mathfrak{H}$. In full generality, it does not have to be a linear, or even algebraic operator, since there are such things as nonlinear and dispersive electromagnetic media. Thus, in its general form it might be more on the order of a nonlinear, integral operator. Hence, in order to make $C$ a linear isomorphism, one is implicitly assuming that the medium is linear and dispersive.

---

([1]) The reason for the more modern terminology "excitation" (cf., Hehl and Obukhov [**8**]), rather than "displacement" or "induction" is that one thinks of $\mathfrak{H}$ as a response of the medium to the imposition of the field strength $F$, which means that we are also ignoring the possibility that the medium has a non-zero conductivity that would imply an induced current, in addition to the formation of electric and magnetic dipoles.

If $C$ is linear and non-dispersive then one can then associate $C$ with a bilinear functional on 2-forms in two ways:

$$C(F, G) = G(C(F)) \text{ or } F(C(G)). \tag{4.1}$$

These do not, however, have to be same functional, since that would be equivalent to the symmetry of the functional $C$. For instance, if we choose the first definition in (4.1) then:

$$C(G, F) = F(C(G)), \tag{4.2}$$

which is the other definition.

Of course, one can always polarize the functional:

$$C = C_+ + C_-, \tag{4.3}$$

where:

$$C_\pm(F, G) = \tfrac{1}{2}[C(F, G) \pm C(G, F)]. \tag{4.4}$$

The anti-symmetric part of $C$ is called its *skewon* part.

If one expresses $F$ in space-time form as:

$$F = \theta^0 \wedge E - \#_s \mathbf{B} \tag{4.5}$$

then the bilinear pairing $F(\mathfrak{H})$ will take the form:

$$F(\mathfrak{H}) = C_+(F, F) = \tilde{\varepsilon}(E, E) - \tilde{\mu}(\mathbf{B}, \mathbf{B}) = \tilde{\varepsilon}^{ij} E_i E_j - \tilde{\mu}_{ij} B^i B^j \ . \tag{4.6}$$

In the simplest case of spatial isotropy, in addition to the aforementioned restrictions, one will have:

$$\tilde{\varepsilon}^{ij} = \varepsilon\, \delta^{ij}, \qquad \tilde{\mu}_{ij} = \frac{1}{\mu}\delta_{ij}, \tag{4.7}$$

and one will get the usual electromagnetic field Lagrangian density:

$$F(\mathfrak{H}) = \varepsilon E^2 - \frac{1}{\mu}\mathbf{B}^2. \tag{4.8}$$

Thus, the Lagrangian density of an electromagnetic field in a medium can be obtained from:

$$F(\mathfrak{H}) = F(C(F)) = C(F, F) = C_+(F, F). \tag{4.9}$$

One sees that the skewon part of $C$ will play no role as far as the field Lagrangian is concerned.

As for the symmetric part $C_+$, one must recall that the volume element $V$ also defines a symmetric bilinear functional by way of:

$$V(F, G) = (F \wedge G)(\mathbf{V}) = \frac{1}{4!} \varepsilon^{\kappa\lambda\mu\nu} F_{\kappa\lambda} F_{\mu\nu}, \tag{4.10}$$

where $\mathbf{V} \in \Lambda_4(M)$ is reciprocal to $V$ (so $V(\mathbf{V}) = 1$).

Hence, one might also decompose $C_+$ into:

$$C_+ = C_0 + \lambda\, V, \tag{4.11}$$

in which $C_0$ becomes the *principal part of C*, while $\lambda\, V$ is the *axion* part.

Typically, (linear, non-dispersive) electromagnetic media tend to be associated with constitutive laws that have only a principal part. The appearance of a skewon part is most often associated with optical phenomena such as the Faraday rotation of the plane of polarization and optical activity [15]. To date, the existence of an axion part has been established more speculatively than empirically (see, however, [16]). Recall that $V(F, F)$ would vanish for any decomposable $F$, which, as we said before, includes many of the most common examples of electromagnetic fields, including wave-like fields. Hence, the axion part of $C$ will not affect the study of electromagnetic waves.

In order to relate all of this to the Hodge * operator, one needs only to define a purely geometric version of the electromagnetic constitutive law $C$ that amounts to "raising both indices of $F$" using the Lorentzian metric $g$:

$$C = (i_g \wedge i_g)^{-1}, \tag{4.12}$$

in which $i_g : T(M) \to T^*(M)$, $\mathbf{v} \mapsto i_g \mathbf{v}$, where $(i_g \mathbf{v})(\mathbf{w}) = g(\mathbf{v}, \mathbf{w})$, so $i_g \mathbf{v} = (g_{\mu\nu} v^\nu)\, \theta^\mu$.

If $(U, x^\mu)$ is a local coordinate chart on $M$, and one looks at components with respect to the natural frame and coframe field then $C$ will have the components:

$$C^{\kappa\lambda\mu\nu} = \tfrac{1}{2}(g^{\kappa\mu} g^{\lambda\nu} - g^{\kappa\nu} g^{\lambda\mu}). \tag{4.13}$$

Thus, the components $F_{\mu\nu}$ will go to the corresponding components $F^{\mu\nu}$ that one conventionally obtains by raising the indices using the metric.

The Hodge * operator is then obtained from the composition $* = \# \cdot C$; that is, it is the map $\kappa$ that is associated with such a $C$.

One can also express the matrix of * in terms of a basis $\{E^I, I = 1, \ldots, 6\}$ for $\Lambda^2$ (see Part I). When the basis is orthonormal for $g$ – in the sense that it is defined by starting with a Lorentzian frame on the cotangent spaces – the matrix of $C$ will become:

$$[C_{IJ}] = \begin{bmatrix} \delta_{ij} & 0 \\ \hline 0 & -\delta_{ij} \end{bmatrix}. \tag{4.14}$$

Thus, as a constitutive law, the Hodge * is linear, isotropic, homogeneous, dispersionless, and has vanishing skewon and axion parts.

The composition of this with the matrix for # then gives the matrix for * in the form:

$$[*^I_J] = [V^{IK}][C_{KJ}] = \begin{bmatrix} 0 & \delta^{ij} \\ \delta^{ij} & 0 \end{bmatrix} \begin{bmatrix} \delta_{ij} & 0 \\ 0 & -\delta_{ij} \end{bmatrix} = \begin{bmatrix} 0 & -\delta^i_j \\ \delta^i_j & 0 \end{bmatrix}. \tag{4.15}$$

Of course, in a non-Lorentzian frame, this will be more involved.

One sees immediately that this matrix has the property of the * operator for a Lorentzian metric that:

$$*^2 = -I. \tag{4.16}$$

This property implies that * defines what one calls an *almost-complex* structure on the vector bundle $\Lambda^2$. As we shall see, it allows one to then define a complex structure on the fibers, which makes them complex-isomorphic to $\mathbb{C}^3$, as well as a complex orthogonal structure that relates very intimately to the propagation of electromagnetic waves for the simplest constitutive law.

One finds (cf., the author's analysis [**17**]) that not every physically-meaningful electromagnetic constitutive law will admit a $\kappa$ that behaves like *. In particular, an anisotropic optical medium will typically not produce a $\kappa$ map that defines an almost-complex structure, although the simplest constitutive law (i.e., linear, isotropic, dispersionless) will imply such a structure, up to a constant multiplier, in the sense that if the matrix of $C$ takes the form:

$$[C_{IJ}] = \begin{bmatrix} \varepsilon \delta_{ij} & 0 \\ 0 & -(1/\mu)\delta_{ij} \end{bmatrix} \tag{4.17}$$

then the matrix of $\kappa$ will take the form:

$$[\kappa^I_J] = \begin{bmatrix} 0 & -(1/\mu)\delta^i_j \\ \varepsilon \delta^i_j & 0 \end{bmatrix}, \tag{4.18}$$

which has the property that:

$$[\kappa^I_J]^2 = -\frac{\varepsilon}{\mu} I. \tag{4.19}$$

The square root of the factor $\varepsilon/\mu$ is sometimes referred to as the *impedance* of spacetime (cf., Hehl and Obukhov [**8**]).

Of course, the latter constitutive law is also the one that gives the Lorentzian dispersion law that defines the usual light cones of relativity. We shall return to discuss the concept of almost-complex structure and the reduction to light cones at several points later in this article.

*b. Constitutive laws and line geometry.* Now, let us return to the world of $M = \mathbb{R}^4$, and, more to the point, lines in $\mathbb{R}P^3$.

A linear, dispersionless, constitutive law $C: \Lambda^2 \to \Lambda_2$ (or really, its inverse) defines a correlation between the associated projective spaces $P\Lambda_2$ and $P\Lambda^2$. Thus, any point of $P\Lambda_2$ will correspond to a unique point in $P\Lambda^2$, which will then represent a hyperplane in $P\Lambda_2$.

If the bilinear functional $C$ is symmetric then one will call that correlation a *polarity*, while if it is anti-symmetric then one will call it a *null polarity*. The symmetric part of $C$ also defines a quadric in $P\Lambda^2$:

$$C_+(F, F) = 0. \tag{4.20}$$

Since we already have a quadric in $P\Lambda^2$ that is defined by the volume element (viz., the Klein quadric), we will focus on the principal part and define the quadric:

$$C_0(F, F) = 0. \tag{4.21}$$

This allows us to define the subset $\mathcal{J}$ of $\mathcal{K}$ that consists of its points that lie on the quadric (4.21). Thus, it will consist of all $F$ such that:

$$V(F, F) = C_0(F, F) = 0. \tag{4.22}$$

We compose the linear isomorphism $C$ with the linear isomorphism $\#: \Lambda_2 \to \Lambda^2$ that is defined by the volume element $V$ and produce a linear isomorphism $\kappa = \# \cdot C : \Lambda^2 \to \Lambda^2$. As mentioned above, this map is the starting point for pre-metric electromagnetism, since it should substitute for the Hodge * operator that usually comes from the Minkowski metric $\eta_{\mu\nu}$ or its Lorentzian generalizations.

In the case of the constitutive law that corresponds to *, the quadric hypersurface is defined by starting with the scalar product of 2-forms:

$$(F, G) = (F \wedge *G)(\mathbf{V}), \tag{4.23}$$

and defining its associated quadratic form:

$$Q_*[F] = (F, F). \tag{4.24}$$

This time, we call the subset of $\mathcal{K}$ that lies on this quadric $\mathcal{I}$ (for *isotropic*). Since:

$$(*F, *G) = - (F, G), \tag{4.25}$$

due to self-adjointness, and the vanishing of one side of this equation will imply the vanishing of the other, the map * will have the property that it takes points of $\mathcal{I}$ to other such points.

Since *C* defines map from P$\Lambda^2$ to P$\Lambda_2$, one must note that when space is four-dimensional both projective spaces will describe lines in $\mathbb{R}P^3$ under the Plücker-Klein embedding. Thus, *C* will map $\mathcal{L}(3)$ to itself.

If we let $\mathfrak{H} = C_0(F)$ then the quadric (4.21) can also be expressed in the form:

$$F(\mathfrak{H}) = 0, \qquad (4.26)$$

and a comparison of this with (4.9) shows that the expression $C_0(F, F)$ is basically the Lagrangian density of the electromagnetic field *F* when the medium is governed by a constitutive law *C* with no axion part, or when the field *F* is decomposable.

Let $\mathbf{e}_\mu$ be a basis for $\mathbb{R}^4$ and let $\mathbf{E}_I$ be a basis for $\Lambda_2$, as usual, and let $\theta^\mu$ and $E^I$ be the corresponding reciprocal bases on $\mathbb{R}^{4*}$ and $\Lambda^2$, resp. Hence, one can express the linear, non-dispersive constitutive law *C* as a doubly-contravariant tensor:

$$C = C^{IJ}\, \mathbf{E}_I \otimes \mathbf{E}_J. \qquad C^{IJ} = \left[\begin{array}{c|c} \varepsilon^{ij} & \alpha^i_{\,j} \\ \hline \beta^i_{\,j} & -\tilde{\mu}_{ij} \end{array}\right]. \qquad (4.27)$$

The notation employed suggests that the matrix $\varepsilon^{ij}$ should represent the dielectric tensor of the medium, $\tilde{\mu}_{ij}$ should represent the inverse of the magnetic permeability, and the cross-coupling matrices $\alpha^i_{\,j}$ and $\beta^i_{\,j}$ should represent the effects of such as things as Faraday rotations and optical activity. Typically, the matrices $\varepsilon$ and $\mu$ are assumed to be symmetric, so the only contribution to the skewon part of *C* would have to come from $\alpha$ and $\beta$.

The components $C^{IJ}$, which have no specific symmetry in their indices, can then be decomposed into a sum:

$$C^{IJ} = C_0^{IJ} + C_-^{IJ} + \lambda V^{IJ}. \qquad (4.28)$$

with:

$$C_0^{IJ} = \left[\begin{array}{c|c} \varepsilon & \gamma_+ - \lambda I \\ \hline \gamma_+^T - \lambda I & -\tilde{\mu} \end{array}\right], \qquad C_-^{IJ} = \left[\begin{array}{c|c} 0 & \gamma_- \\ \hline -\gamma_-^T & 0 \end{array}\right], \qquad \lambda = \tfrac{1}{6}\operatorname{Tr}\kappa, \qquad (4.29)$$

in which we have defined:

$$\gamma_\pm = \tfrac{1}{2}(\alpha \pm \beta^T). \qquad (4.30)$$

When the matrix of *C* is composed with the matrix for #, that will give:

$$[\kappa] = [\#][C] = \left[\begin{array}{c|c} \beta^i_{\,j} & -\tilde{\mu}_{ij} \\ \hline \varepsilon^{ij} & \alpha^i_{\,j} \end{array}\right]. \qquad (4.31)$$

One can also form:

$$[\kappa_0] = [\#][C_0] = \left[\begin{array}{c|c} \gamma^T & -\tilde{\mu} \\ \hline \varepsilon & \gamma \end{array}\right]. \qquad (4.32)$$

One of the properties of $C_+$ or $C_0$ that it inherits from its symmetry as a bilinear functional is that the associated linear map $\kappa_+$ or $\kappa_0$ is *self-adjoint;* since $\kappa_0$ does not involve the bilinear functional that comes $V$, we choose to use it and assert:

**Theorem:**

*For any $F, G \in \Lambda^2$, one must always have:*

$$\kappa_0(F) \wedge G = F \wedge \kappa_0(G).$$

Proof:

$$\kappa_0(F) \wedge G = C_0(F, G)\, V = C_0(G, F)\, V = F \wedge \kappa_0(G).$$

The question then arises of the extent to which $\kappa_0$ maps $\mathcal{J}$ to itself, just as $*$ mapped $\mathcal{I}$ to itself. Now $F \in \mathcal{J}$ iff:

$$F \wedge F = F \wedge \kappa_0(F) = 0. \qquad (4.33)$$

Thus, $\kappa_0(F)$ will belong to $\mathcal{J}$ iff:

$$\kappa_0(F) \wedge \kappa_0(F) = \kappa_0(F) \wedge \kappa_0^2(F) = 0. \qquad (4.34)$$

Now, from self-adjointness, one will have:

$$\kappa_0(F) \wedge \kappa_0(F) = F \wedge \kappa_0^2(F), \quad \kappa_0(F) \wedge \kappa_0^2(F) = F \wedge \kappa_0^3(F),$$

so the vanishing of these expressions will be closely related to the behavior of the powers of $\kappa_0$. In particular, it would be sufficient that $\kappa_0$ had either property:

$$\kappa_0^2 = \zeta^2\, I \qquad \text{or} \qquad \kappa_0^2 = \zeta^2\, \kappa_0. \qquad (4.35)$$

The first of these is the closest analogue of the condition $*^2 = -I$ for the Hodge star operator, for which $\zeta = i$, and also includes the case of linear, isotropic media, for which $\zeta^2 = -\varepsilon/\mu$.

If one forms the matrix of $\kappa_0^2$ using (4.32) then one will get:

$$[\kappa_0^2] = \left[\begin{array}{cc} \varepsilon^2 - \gamma\gamma^T & \varepsilon\gamma - \gamma\mu \\ \gamma^T\varepsilon - \mu\gamma^T & \mu^2 + \gamma^T\gamma \end{array}\right]. \qquad (4.36)$$

Thus, the first condition in (4.35) becomes the three conditions:

$$\varepsilon^2 - \gamma \gamma^T = \tilde{\varepsilon}^2 - \gamma \gamma^T = \zeta^2 I, \qquad \varepsilon \gamma - \gamma \mu = 0 \qquad (4.37)$$

on the submatrices.

One then sees how restricting these conditions are on the electromagnetic properties of the space in question, since even when $\gamma = 0$, one must have $\varepsilon$ and $\mu$ matrices that square to $I$, which generally means that the medium is isotropic, and even then one must adjust the definition of the dielectric constant and magnetic permeability in order to justify making them equal to each other. One should also not that these conditions can be satisfied by media that are what Lindell [**18**] calls *bi-isotropic*, which differ from isotropic by allowing $\gamma$ to be proportional to the identity matrix, as well. Further discussion of how one gets from $\kappa_0$ to * is found in the author's paper [**17**].

An important example in which these conditions are not satisfied is the case of a linear, anisotropic, optical medium. To say that it is optical means that it is magnetically isotropic, and generally homogeneous, so $\tilde{\mu}_{ij} = 1/\mu \, \delta_{ij}$. Along with setting $\gamma = 0$, one usually moves to a principal frame for the symmetric matrix $\varepsilon^{ij}$, in which it becomes diagonal. Thus:

$$C_0^{IJ} = \begin{bmatrix} \varepsilon_x & 0 & 0 & \\ 0 & \varepsilon_y & 0 & 0 \\ 0 & 0 & \varepsilon_z & \\ \hline 0 & & & -(1/\mu)I \end{bmatrix}. \qquad (4.38)$$

One sees from (4.34) that even expecting $\kappa_0$ to take the Klein quadric $\mathcal{K}$ to itself is debatable enough, to begin with. Hence, $\kappa_0$ does have to take decomposable 2-forms to other of their kind, and in turn, planes through the origin in $\mathbb{R}^4$ and lines in $\mathbb{RP}^3$.

The introduction of a linear, dispersionless constitutive law into the spacetime manifold allows one to define a quadratic line complex by way of its symmetric part, namely:

$$V(F, F) = 0, \quad C_+(F, F) = 0. \qquad (4.39)$$

(Here, we are describing lines as axes, in the language of Plücker.)

If we introduce scalar products that are defined by $V$ and $C$:

$$\langle F, G \rangle \equiv V(F, G), \qquad (F, G) \equiv C_+(F, G) \qquad (4.40)$$

then we can express the complex in an even more manifestly geometric form:

$$\langle F, F \rangle = 0, \quad (F, F) = 0. \qquad (4.41)$$

We shall refer to any $F$ that belongs to this quadratic complex as *isotropic*.

For the case of the Hodge * isomorphism, one can go further, since the almost-complex structure that it defines allows one to also define a complex structure on the fibers of $\Lambda^2$ by using * to represent $i$:

$$iF \equiv *F. \quad (4.42)$$

One then defines the multiplication of 2-forms by complex scalars by:

$$(\alpha + i\beta)\, F \equiv \alpha F + \beta *F. \quad (4.43)$$

One finds that the basis $\{E^i, *E^i, i = 1, 2, 3\}$ for $\Lambda^2$ that we have defined in Part I does in fact have the property that $*(*E^i) = -E^i$, which allows us to use $\{E^i, i = 1, 2, 3\}$ as a *complex* 3-frame on each fiber of $\Lambda^2$ by way of $\mathbb{C}^3 \to \Lambda^2$, $Z_i \mapsto Z_i E^i$. Thus, any 2-form $F$ can be represented as:

$$F = E_i E^i + B_i *E^i = (E_i + iB_i)\, E^i. \quad (4.44)$$

The representation of the electromagnetic field strengths as a complex 3-vector goes back to Riemann [19], and was revisited by Silberstein [20], Majorana [21], and Oppenheimer [22]. It also finds a place in the complex formulation of general relativity, which independently derives a complex formulation of the Einstein field equations for gravitation that reduce to a generalized Maxwellian form for weak fields (see the author's paper [23] and the references cited therein).

The * operator also allows one to define the second scalar product in the form:

$$(F, G) \equiv \langle F, *G \rangle, \quad (4.45)$$

and, in fact, one can combine the two scalar products into one complex scalar product:

$$\langle F, G \rangle_{\mathbb{C}} \equiv (F, G) + i \langle F, G \rangle. \quad (4.46)$$

The basis $\{E^i, i = 1, 2, 3\}$ then becomes complex-orthonormal for this scalar product:

$$\langle E^i, E^j \rangle_{\mathbb{C}} = (E^i, E^j) + i \langle E^i, E^j \rangle = \delta^{ij}. \quad (4.47)$$

One can then say that the quadratic line complex $\mathcal{I}$ on $\mathbb{R}P^3$ is defined by all $F$ such that:

$$\langle F, F \rangle_{\mathbb{C}} = 0, \quad (4.48)$$

which we can still refer to as isotropic lines.

Since this condition is homogeneous, it will not only define a complex quadric on $\mathbb{C}^3$, in the form of the sphere of radius zero, but also a complex quadric on $\mathbb{C}P^2$, which consists of all complex lines through the origin of $\mathbb{C}^3$. Since $\mathbb{C}P^2$ is four-dimensional as a real manifold, the level surface of a complex function on it will be two-dimensional as a real manifold. Thus, the specialization that is associated with the constitutive law that represents the Hodge * operator brings one into the realm of complex projective geometry, which would take us beyond the scope of the present study.

*c. Electromagnetic dispersion laws.* The dispersion law that goes with the almost-complex structure * on $\Lambda^2$ (viz., the light-cone) might be most commonly-discussed one at the elementary level, but it is not the only one that is possible, nor is it the only one that is generally recognized as important to physics. Typically, dispersion laws come out of optical applications, but one can also consider them in the study of electromagnetic waves in plasmas, and even massive (i.e., Klein-Gordon) waves, which have inhomogeneous dispersion laws that resemble some of the plasma laws.

Customarily, the way that one gets from the imposition of an electromagnetic constitutive law on the spacetime manifold to an associated dispersion law for electromagnetic waves in that medium is by starting with the most general field equations for electromagnetism and restricting the class of solutions to ones that are "wave-like." Of course, the latter notion is quite open-ended, and it is only when one is dealing with a linear constitutive law that it is sufficient to assume a plane-wave form for the wave-like fields, since in the absence of superposition the dispersion law for a more complex wave-form that is built up from plane waves by Fourier analysis might very well be different from the dispersion law for the plane-waves themselves. (This is closely related to the way that the frequency of a nonlinear oscillator − e.g., a physical pendulum − might depend upon its amplitude, unlike the situation for a linear oscillator.)

For the purposes of linear, dispersionless media, the most direct route to getting a dispersion law from a constitutive law is by looking at the dispersion law for plane waves.

One first expresses the electromagnetic field strength 2-form $F$ as the exterior derivative $dA$ of an electromagnetic potential 1-form $A$ (which might be possible only locally) so the pre-metric Maxwell equations (*in vacuo*), viz.:

$$F = dA, \qquad \mathfrak{H} = C(F), \qquad \text{div } \mathfrak{H} = 0, \qquad (4.49)$$

can be combined into a single second-order equation for $A$:

$$\Box_C A = 0, \qquad (4.50)$$

in which the field operator $\Box_C : \Lambda^1 \to \Lambda_1$, takes the form:

$$\Box_C = \text{div} \cdot C \cdot d. \qquad (4.51)$$

If one lets $A$ take the form $e^{i\phi} a$, where $\phi$ is a phase function on the spacetime manifold and $da = 0$, then $F$ will take the form:

$$F = i\,(k \wedge A), \tag{4.52}$$

in which $k = d\phi$.

With $k = \omega\,dt + k_i\,dx^i$, one will then get:

$$F = i\,(dt \wedge \omega A + k_s \wedge A), \tag{4.53}$$

so one can associate the electric and magnetic field strengths by way of:

$$E = i\omega A, \qquad B = i k_s \wedge A. \tag{4.54}$$

For our purposes, only the $k \wedge A$ part of the expression (4.52) will be meaningful. One then sees that it will take the form of the fundamental 2-form for the instantaneous wave surfaces in space as long as the 1-form $a$ lies in the plane of $k$ and $dt$, without being collinear with $k$; in particular, since $k = \omega\,dt + k_i\,dx^i$, it can be purely spatial.

The line $[k, A]$ in $\mathbb{R}P^3$ that is represented (as an axis) by the 2-form $F = k \wedge A$ is the intersection of the planes $[k]$ and $[A]$. If $A$ is purely spatial in $\mathbb{R}^{4*}$ then $[A]$ will be a point at infinity, and since $k$ has the homogeneous coordinates $(\omega, k_i)$, the point $[k]$ will be finite, but not equal to the point with the inhomogeneous coordinates $(0, 0, 0)$, which corresponds to $dt$. Hence, the line $[k, A]$ will be a line to infinity that does not pass through the origin (of the inhomogeneous coordinate system).

With this same substitution, the field operator $\Box_C$ will go to (a non-zero scalar multiple of) its Fourier transform (i.e., its symbol):

$$\sigma(\Box_C, k) = i_k \cdot C \cdot e_k, \tag{4.55}$$

in which the operators $i_k$ and $e_k$ are defined by:

$$i_k : \Lambda_2 \to \Lambda_1,\; \mathbf{a} \wedge \mathbf{b} \mapsto k(\mathbf{a})\,\mathbf{b} - k(\mathbf{b})\,\mathbf{a},$$

$$e_k : \Lambda^1 \to \Lambda^2,\; a \mapsto k \wedge a.$$

That is, $i_k$ represents interior multiplication by $k$, while $e_k$ represents exterior multiplication by $k$

If we look at components with respect to some frame field on $\mathbb{R}^4$ and its reciprocal coframe field then the linear, algebraic operator $\sigma(\Box_C, k)$, which we will write as $C(k)$, will have the matrix:

$$C(k)^{\mu\nu} = C^{\kappa\mu\lambda\nu} k_\kappa k_\lambda. \tag{4.56}$$

Thus, $C(k)^{\mu\nu}$ is a homogeneous, quadratic polynomial in the components of $k$. If one confines oneself to the symmetric part $C_+$ of $C$ then $C_+(k)^{\mu\nu}$ will also be a symmetric matrix.

It is tempting to say that $C_+(k)$ then defines a cone in the cotangent space to each $k$, but one finds that it can never be invertible (i.e., non-degenerate) as a linear operator, since the kernel of $e_k$ is the line spanned by $k$ and the image of $i_k$ is the hyperplane that is annihilated by $k$. Hence, one must restrict $C_+(k)$ to a hyperplane in each cotangent space that is transverse to the line $[k]$ if one is to have any hope of making $C_+(k)$ invertible, at least for generic values of $k$. One then converts to an adapted coframe $\bar{\theta}^0 = k$, $\bar{\theta}^a$, $a = 1$, 2, 3 for which the $\bar{\theta}^a$ span the transverse hyperplane, and considers the 3×3 submatrix $C(k)^{ab}$.

Whether or not $C_+(k)^{ab}$ is invertible will now depend upon $k$, and the equation:

$$Q[k] \equiv \det C_+(k)^{ab} = 0 \tag{4.57}$$

will define a homogeneous, sextic polynomial equation in $k$, although typically it will reduce to a homogeneous quartic. Hence, it will define a quartic cone in each cotangent space.

The determinant in question can also be expressed as a 4-form on the spacetime manifold that is usually called the *Tamm-Rubilar tensor field*, since it was first proposed by Tamm [**24**] and then more recently refined by Rubilar [**25**]. For a discussion of it, one can consult Hehl and Obukhov [**8**].

There is then a succession of increasingly specialized cases of $Q[k]$ that lead back to the Lorentzian metric, which one can regard as being rooted in the dispersion law for the propagation of electromagnetic waves in spacetime. The first specialization is to quartic polynomials that factor into the product $g_1(k, k)\, g_2(k, k)$ of two quadratic polynomials of Lorentzian type – i.e., signature type $(+1, -1, -1, -1)$ – which is the *bimetric* case. A further degeneracy can make the quartic polynomial become the square $g(k, k)^2$ of a single quadratic polynomial of Lorentzian type. This is the point at which pre-metric electromagnetism rejoins the usual relativistic formulation, although since the constitutive laws that will lead to the Lorentzian case of a dispersion law are only one specialized possibility that basically pertains to the classical vacuum, one sees that the scope of electromagnetism is essentially larger than that of the Lorentzian geometry that apparently explains the presence of gravitation in spacetime – at least, à la Einstein. One also sees that, in one sense, the theories of electromagnetism and gravitation have been "unified" – or at least, logically connected – all along; indeed, Einstein himself started out talking about electromagnetism and ended up talking about gravitation.

*d. Dispersion laws and line geometry.* One can now see that the process by which one goes from a linear, dispersionless constitutive law (or at least, its symmetric part) to a dispersion law for plane electromagnetic waves is essentially the same as the process by which one goes from a quadratic line complex on $\mathbb{RP}^3$ to its singularity surface, since the matrix $C_+(k)^{ab}$ plays the same role that $C(X_0)^{ij}$ did above. Of course, we should once more point out that since we are dealing with 1-forms and 2-forms here, we are using what Plücker referred to as the representation of line in $\mathbb{RP}^3$ as axes; i.e., the intersection of planes. Hence, $C_+(k)^{ab}$ will define a curve enveloped by lines in the plane that is defined by *k*, and the degeneracy in its determinant will make that curve degenerate to two distinct points.

In general, the quartic surface that is defined in $\mathbb{RP}^3$ by $Q[k]$ is a Kummer surface, but in the linear, dispersionless, optical case, it reduces to a *Fresnel (or wave) surface* [**6, 26, 27, 28**]. In that case, the constitutive law takes the simple form:

$$[C]^{IJ} = \begin{bmatrix} \varepsilon_x & 0 & 0 & & \\ 0 & \varepsilon_y & 0 & & 0 \\ 0 & 0 & \varepsilon_y & & \\ \hline & & & & \frac{1}{\mu}\delta_{ij} \\ & 0 & & & \end{bmatrix}, \tag{4.58}$$

in which one has chosen a principal frame for the dielectric tensor $\varepsilon^{ij}$.

One notes that this form of constitutive law is not only symmetric, so it has a vanishing skewon part, but also has no axion part, due to the zeroes in the off-diagonal sub-matrices. Hence, it coincides with its principal part.

Initially, the equation of the *Fresnel normal surface* takes the form (cf., Born and Wolf [**26**]):

$$\frac{K_x^2}{n^2 - \mu\varepsilon_x} + \frac{K_y^2}{n^2 - \mu\varepsilon_y} + \frac{K_z^2}{n^2 - \mu\varepsilon_z} = \frac{1}{n^2}, \tag{4.59}$$

in which the actual speed of propagation of a plane wave through the medium is expressed by $v_p = c / n$ and $K_x$, $K_y$, $K_z$ represent the inhomogeneous coordinates for the point of $\mathbb{RP}^{3*}$ that is associated with the covector *k*.

This equation can be converted further by introducing three *principal speeds of propagation:*

$$v_i = \frac{1}{\sqrt{\mu\varepsilon_i}}, \quad i = x, y, z, \tag{4.60}$$

and one gets:

$$\frac{K_x^2}{v_p^2 - v_x^2} + \frac{K_y^2}{v_p^2 - v_y^2} + \frac{K_z^2}{v_p^2 - v_z^2} = 0. \tag{4.61}$$

By clearing the fractions, one sees that this equation can be regarded as a quadratic equation in $v_p^2$; i.e., a quartic in $v_p$. Thus, for a given direction $K_i$ of the wave normal, there will generally be *two distinct* values for $v_p^2$; the fact that $v_p^2$ has two square roots says only that the wave can propagate forward or backward. The actual value that is chosen by the wave then depends upon its state of polarization.

One refers to this double-valuedness of the propagation speed as either *double refraction* or *birefringence*. It is most commonly observed in calcite crystals, although it also defines the way that jewelers can distinguish diamonds, which are birefringent, from glass, which is not.

There is a dual construction to the Fresnel normal surface that pertains to the directions of the *rays* that correspond to the propagation of the waves. The direction $\mathbf{t} = (t^x, t^y, t^z)$ of the ray is determined by the flow of energy in the wave (i.e., the Poynting vector), and does not have to coincide with the normal to the wave surface. This time, one has a ray velocity $v_r = c / n_r$ and a corresponding index of refraction $n_r$, and if one denotes the inverses of the principal speeds by $n_x$, $n_y$, $n_z$, which one then thinks of as *principal indices of refraction*, then the equation of the Fresnel ray surface will take the form:

$$\frac{(t^x)^2}{n_r^2 - n_x^2} + \frac{(t^y)^2}{n_r^2 - n_y^2} + \frac{(t^z)^2}{n_r^2 - n_z^2} = 0. \qquad (4.62)$$

Once again, this gives a quadratic equation in $n_r^2$ – or $v_r^2$, for that matter – so any given ray direction is associated with two ray velocities.

The Fresnel wave surface – in either of its forms – does not actually define a differentiable manifold, since it intersects itself, and thus has singular points at which the tangent lines generate cones. In fact, the Fresnel wave surface is a degenerate case of a Kummer surface for which the sixteen singular points are distributed over four tangent planes with four points on each, which is why it is sometimes referred to as a "tetrahedroid" (cf., Hudson [**6**]). One can also define a quadratic line complex that one calls the "Painvin complex" that produces the Fresnel wave surface (4.61) as its singularity surface directly (see Hudson [**6**]).

The singular points are associated with an important optical phenomenon, in the form of *conical refraction*. It can take the form of *external* or *internal* conical refraction depending upon whether one is looking at singularities of the normal or ray surface, respectively. Basically, along a singular axis, which goes through the origin and a singular point, a given wave normal will be associated with a cone of ray directions, which means that the correspondence between them is not invertible. Dually, the ray surface has the property that some tangent planes contact it at an infinitude of points, and thus envelop a cone. The cone of rays that is associated with a singular wave normal leads to internal conical refraction, while the cone of wave normals that is associated with a singular ray leads to external conical refraction. The experimental verification of the phenomenon of internal conical refraction, which had been predicted by Hamilton in 1832 on the basis of his theory of geometrical optics ([1]), was regarded as a significant triumph of that theory.

The association of a quadratic dispersion law with the constitutive law that gives the Hodge * operator can be obtained more directly than by first going to the quartic and reducing to a quadratic. Let $k = k_0 \theta^0 + k_s$, and let $a$ be purely spatial, so:

$$F = k \wedge a = k_0 \theta^0 \wedge a + k_s \wedge a. \qquad (4.63)$$

---

([1]) Interestingly, Hamilton's theory of geometrical optics [**29**], which still came down to his celebrated system of ordinary differential equations, historically preceded his theory of mechanics that was based upon the same basic considerations.

In order for $F$ to be isotropic, one must have:

$$0 = (F, F) = k_0^2 (\theta^0 \wedge a, \theta^0 \wedge a) + (k_s \wedge a, k_s \wedge a). \tag{4.64}$$

We can now use the linear isomorphism $e_0 : \mathbb{R}^{3*} \to \Lambda_+^2$, $a \mapsto \theta^0 \wedge a$ to define a scalar product on spatial covectors:

$$(a, b) \equiv (\theta^0 \wedge a, \theta^0 \wedge b), \tag{4.65}$$

which will make:

$$(\theta^0 \wedge a, \theta^0 \wedge a) = \| a \|^2. \tag{4.66}$$

One can also use the linear isomorphism $* : \Lambda_+^2 \to \Lambda_-^2$ to define another scalar product on $\mathbb{R}^{3*}$. If one lets the spatial 2-forms $a \wedge b$ and $c \wedge d$ take the forms $*(\theta^0 \wedge u)$ and $*(\theta^0 \wedge v)$, respectively, then one can set:

$$(a \wedge b, c \wedge d) = (*(\theta^0 \wedge u), *(\theta^0 \wedge v)) = - (\theta^0 \wedge u, \theta^0 \wedge v) = - (u, v). \tag{4.67}$$

One can also use a basic identity of the vector cross product (suitably adapted to the exterior product):

$$(a \wedge b, c \wedge d) = - (a, c)(b, d) + (a, b)(c, d), \tag{4.68}$$

which makes:

$$(k_s \wedge a, k_s \wedge a) = [k_0^2 - (k_s, k_s)] \| a \|^2 + (k_s, a)^2. \tag{4.69}$$

This allows us to say:

**Theorem:**

*The line in $\mathbb{R}P^3$ that is described by the 2-form $k \wedge a$ is isotropic if:*

$$0 = k_0^2 - (k_s, k_s), \qquad 0 = (k_s, a). \tag{4.70}$$

Since the same 2-form $k \wedge a$ can be spanned by an infinitude of other pairs of covectors $k$ and $a$, one cannot hope to assert the validity of the converse of this theorem. One sees this from the fact that the right-hand side of (4.69) can vanish for some spatial $a$ when $k_0^2 - (k_s, k_s) < 0$.

The first equation in (4.70) has the form of a Lorentzian light cone, which is defined by the most elementary dispersion law for electromagnetic waves when one sets $k_0 = \omega / c$. The fact that it is homogeneous in $k$ implies that any other metric that is conformal to the Minkowski space metric would give the same light cone, so one cannot say that the dispersion law defines a unique Lorentzian metric, but only a conformal class of them.

Since the scalar product on $\mathbb{R}^{3*}$ is Euclidian, the second equation in (4.70) says that $a$ must be perpendicular to $k_s$, and if one expresses this in component form as:

$$k_i \, a^i = 0 \qquad (4.71)$$

then one will see that this condition is similar to the Fourier transform of the vanishing of a spatial divergence of a spatial 1-form, which then takes the form of the Coulomb gauge condition.

**5. Summary.** As we did before in Part I of this series, we first summarize the basic results in the form of a table that correlates the physical notions in one column with the line-geometric ones in the other column.

Table 1. Electromagnetic concepts vs. line-geometric ones

| **Electromagnetism** | **Line geometry** |
|---|---|
| Tangent plane to an instantaneous wave surface | Line in $\mathbb{R}P^3$ (projectivized tangent space) <br> Intersection of planes [k] and [dt] |
| Wave-like electromagnetic field strength $F = k \wedge a$ | Line to infinity [k, a] <br> Intersection of planes [k] and [a] |
| Linear, dispersionless, constitutive law (its symmetric part) | Quadratic line complex on $\mathbb{R}P^3$ |
| Constitutive law for the Hodge * isomorphism | Almost-complex structure on $\Lambda^2$ <br> Complex quadric on $\mathbb{C}P^2$ (the space of isotropic lines) |
| Dispersion law for plane waves in a linear, dispersionless medium | Singularity surface of the quadratic line complex defined by the constitutive law (E.g., Kummer surface) |
| Dispersion law for optical media | Fresnel wave surface |
| Dispersion law for linear, isotropic, homogeneous, dispersionless media | Conformal Lorentz structure on cotangent spaces |

One thing becomes clear from the table: The extension of the methods of line geometry to media that are not linear and dispersionless will require some fundamental re-thinking of the problem. In particular, the constitutive law that is associated with the Heisenberg-Euler one-loop effective field theory of quantum electromagnetism is nonlinear, but homogeneous and dispersionless. It does seem to lead to a quartic dispersion law of the bi-metric form for weak fields, along with the possibility of vacuum birefringence, so one assumes that the basic methodology is still useful.

---

(*) References marked with an asterisk are available in English translation at the author's website: neo-classical-physics.info.

Line geometry and electromagnetism II: wave motion 31

c.  "Second supplement to an essay on the theory of systems of rays," Trans. Roy. Irish Acad. **16**, part 2 (1831), 93-125.
d.  "Third supplement to an essay on the theory of systems of rays," Trans. Roy. Irish Acad. **17**, part 1 (1837), 1-144.